\documentclass[draft]{ws-procs961x669}

\usepackage[utf8]{inputenc}
\usepackage{amsmath}
\usepackage{amssymb}
\usepackage{bbm}
\usepackage{bm}
\usepackage{color}
\numberwithin{equation}{section}
\numberwithin{figure}{section}
\usepackage{physics}
\usepackage{graphicx}
\usepackage{epstopdf}
\usepackage{aligned-overset}
\usepackage{hyperref}
\hypersetup{colorlinks=true,linktoc=page,linkcolor=blue,citecolor=red,urlcolor=cyan}

\usepackage[english]{babel}

\newcommand{\hu}{\hspace{0.6cm}}

\newcommand{\dx}[1][x]{{{\rm d} #1}}

\newcommand{\dime}{d}

\usepackage{xparse}

\NewDocumentCommand{\dkpara}{O{k} O{\dime}}{\frac{{\rm d} {#1}^{\parallel}}{(2\pi)^{#2}}}
\NewDocumentCommand{\dkd}{O{k} O{\dime}}{ \frac{{\rm d}^{#2} {#1}}{(2\pi)^{#2}}}
\NewDocumentCommand{\dxd}{O{x} O{\dime}}{ {\rm d}^{#2} {#1} }
\NewDocumentCommand{\ad}{O{} O{}}{ \bigl \langle {#1}\bigr\rangle_{{\rm ad}{#2}} }

 \counterwithout{equation}{section}

\begin{document}

\wstoc{For proceedings contributors: Using World Scientific's\\ WS-procs961x669 document class in \LaTeX2e}
{A. B. Author and C. D. Author}

\title{Strong-field resummed heat kernels and effective actions: inhomogeneous fields}

\author{S.~A.~Franchino-Viñas$^{1,2,3,4}$}

\address{$^1$ Helmholtz-Zentrum Dresden-Rossendorf, Bautzner Landstra{\ss}e 400, 01328 Dresden, Germany}
\address{$^2$ DIME, Universit\`a di Genova, Via all'Opera Pia 15, 16145 Genova, Italy}
\address{$^3$ INFN Sezione di Genova, Via Dodecaneso 33, 16146 Genova, Italy}
\address{$^{4}$
Laboratoire d’Annecy-le-Vieux de Physique Théorique, CNRS – USMB, BP 110 Annecy-le-Vieux,
74941 Annecy, France}

\author{C.~Garc\'ia-P\'erez$^{2,3}$}
\author{F.~D.~Mazzitelli$^{5,6}$}
\address{$^5$ Centro At\'omico Bariloche,  CONICET,
Comisi\'on Nacional de Energ\'\i a At\'omica, R8402AGP Bariloche, Argentina}

\address{$^6$
Instituto Balseiro, Universidad Nacional de Cuyo, R8402AGP Bariloche, Argentina. }

\author{S.~Pla$^{7,8}$}
\address{$^7$ Theoretical Particle Physics and Cosmology, King’s College London, WC2R 2LS, UK}

\address{$^8$ Physik-Department, Technische Universit\"at M\"unchen,
James-Franck-Str., 85748 Garching, Germany}

\author{V.~Vitagliano$^{2,3}$}
\author{U.~Wainstein-Haimovichi$^{9,10}$}
\address{$^9$ Departamento de F\'isica, Facultad de Ciencias Exactas,
Universidad Nacional de La Plata, C.C.\ 67, 1900 La Plata, Argentina.}
\address{$^{10}$ Instituto de F\'isica La Plata (UNLP-CONICET), Diagonal 113 entre 63 y 64, 1900 La Plata, Argentina}

\begin{abstract}
We study the strong-field limit of a theory involving a quantum scalar field coupled to a vector background, which can be either an electromagnetic field or a non-gauge field coupled through the first derivative term.
Our approach consists in obtaining resummed expressions for the associated heat kernels, from which we derive the corresponding resummed effective actions. These results allow us to discuss the effect of pair creation.
Finally, we conjecture that resummations for more general theories should be possible.
\end{abstract}


\bodymatter

\section{Introduction}\label{sec:intro}

Almost ninety years have elapsed since the groundbreaking contributions of Heisenberg and his students, Euler and Kockel~\cite{Euler:1935zz,Heisenberg:1936nmg}, which reshaped our understanding of Quantum Electrodynamics (QED) in the presence of background fields.
Their pioneering works laid the foundation for deriving an effective action that incorporates quantum corrections from fermionic loops and governs the dynamics of the electromagnetic field in the homogeneous field limit.
The importance of these findings remains significant, as they are central to contemporary efforts aimed at the direct detection of the elusive quantum effects of the QED vacuum~\cite{Ahmadiniaz:2022nrv, Gies:2022pla, Macleod:2024jxl}.

These theoretical predictions coincide with the emergence of a new generation of laser facilities, which are steadily pushing the frontier of high intensities: several experiments with previously unimaginable field strengths are currently underway or being tested for feasibility, including some at the European XFEL~\cite{Ahmadiniaz:2024xob, LUXE:2023crk}, LASERIX~\cite{Kraych:2024wwd} and OVAL~\cite{Fan:2017fnd}.

Although considering inhomogeneities in the electromagnetic fields may not be necessary at present due to the limited experimental capabilities, their significance is expected to grow in the near future. Recently, some theoretical tools have been devised to address this scenario, including  the theory of resurgence~\cite{Dunne:2022esi} and resummed heat kernel expansions~\cite{Franchino-Vinas:2023wea}. By {\em resummed expressions}, we mean those  that encapsulate an infinite number of terms appertaining to a particular class, e.g. those which provide the leading order in the strong field regime.

In this manuscript, we will employ heat-kernel techniques to derive  resummed effective actions in a simple quantum  scalar theory with two different classical vector couplings. Closely related results are the resummations valid for rapidly varying curvatures in the so-called covariant perturbation theory~\cite{Barvinsky:1987uw, Barvinsky:1990up,Gorbar:2002pw, Gorbar:2003yt, Franchino-Vinas:2018gzr,Silva:2023lts}, for a large Ricci scalar~\cite{Parker:1984dj,Hu:1984js,Jack:1985mw,Flachi:2014jra, Flachi:2015sva, Flachi:2019btk, Castro:2018iqt}, for couplings involving fields at boundaries~\cite{Franchino-Vinas:2020okl, Edwards:2021cyp, Ahmadiniaz:2022bwy, Franchino-Vinas:2022yka}, for the derivative expansion in QED~\cite{Gusynin:1998bt}, as well as the splitting of heat-kernel contributions into connected and disconnected parts~\cite{Kennedy:1995}. Heat-kernel approaches have also recently been used to study higher-derivative operators~\cite{Barvinsky:2024kgt} and higher-spin fields~\cite{Ferrero:2023xsf}.

Our calculation relies on the key formula linking the Euclidean effective action, $\Gamma_{\rm E}$, to the diagonal of the heat kernel of the operator of quantum fluctuations, $K(x,x;\tau)$ (some examples will be shown in the following sections):
\begin{align}
\label{eq:effective_action}
\Gamma_{\rm E}=- n_s \int \dxd[x][\dime] \int_0^\infty \frac{\dx[\tau]}{\tau}  K(x,x;\tau).
\end{align}
Remarkably, this relation holds for any number of spacetime dimensions $d$ and is applicable regardless of the type of background and quantum fields under consideration (for the latter, the constant $n_s$ is chosen accordingly). Note also that although in our discussion below we will work in flat spacetime, the machinery is particularly suitable for curved spacetime computations.

Using  Eq.~\eqref{eq:effective_action}, in Sec.~\ref{sec:abelian} we will review how the ``powers of the  field strength $F_{\mu\nu}$'' can be resummed.
This was  particularly useful to some of the authors of the present manuscript to prove in Ref.~\citenum{Franchino-Vinas:2023wea} the conjecture proposed by Navarro-Salas and Pla~\cite{Navarro-Salas:2020oew}, which states that, in four spacetime dimensions, the invariants\footnote{$^*F_{\mu\nu}$ is the Hodge dual of the field strength.}
\begin{align}
\mathcal{F}:=F_{\mu\nu}F^{\mu\nu}\quad \text{and}\quad \mathcal{G}:=F_{\mu\nu}{}^*F^{\mu\nu}
\end{align}
appear as an overall ``Euler--Heisenberg prefactor'' in the effective action. In fact, the conjecture turned out to be  true and extendable to a broader context: it holds in any number of spacetime dimensions and is satisfied at the local level of the heat kernel, i.e. without the need to perform integration by parts. In Sec.~\ref{sec:non-gauge}, we will argue that similar resummations should be available for other heat kernels and their corresponding effective actions; to this end, we will present a simple example involving a derivative coupling with a non-gauge vector field. Finally, we will conclude in Sec.~\ref{sec:outlook} with a discussion of potential future research directions.


\section{Resummation for an Abelian gauge background}\label{sec:abelian}
Let us briefly review the results obtained in Ref.~\citenum{Franchino-Vinas:2023wea}.
We focus on a complex, quantum  scalar field $\phi$ interacting with an Abelian background gauge field $A_\mu$, as described by the action
\begin{align}\label{eq:SEM}
 S_{EM}:= \int \dxd[x][\dime] \,\phi^\dagger\, (\nabla- A)^2 \phi.
\end{align}
The background field $A_\mu$ can be arbitrary, provided its field strength is sufficiently large, making a resummation of its powers convenient. In this manuscript, the operator $\nabla$ denotes partial derivatives, indicating that we are operating within flat spacetime. However, we could apply a covariantization argument at the final stages of our derivations to obtain general formulae applicable in curved spacetime.

The operator of quantum fluctuations  related to the expression~\eqref{eq:SEM} can be written as
\begin{align}\label{eq:operatorQ}
    \mathcal{Q}:=- \nabla^2 +2A_\mu(x) \nabla^\mu + V_{EM}(x),
\end{align}
where we have hidden two different contributions in the definition of the potential
\begin{align}\label{eq:VforEM}
 V_{EM}(x):&= \nabla_\mu A^\mu-A_\mu A^\mu.
 \end{align}
Its heat kernel $K_{EM}$, formally defined as $K_{EM}:=e^{-\tau \mathcal{Q}}$, satisfies a parabolic differential equation with a suitable initial condition in the proper time $\tau$,
\begin{align}\label{eq:HK_eq_EM}
 [\partial_\tau +\mathcal{Q}] K_{EM}(x,x';\tau)&=0,
 \\
 \label{eq:HK_initial_condition}
 K_{EM}(x,x',0^+)&= \delta(x-x').
 \end{align}

In principle, any gauge condition can be chosen for the background electromagnetic field $A_\mu$; however, it is convenient to adopt the Schwinger--Fock condition,
\begin{align}
 (x-x')^\mu A_\mu(x)=0\,.
\end{align}
This choice allows us to expand the gauge field in terms of (derivatives of) its field strength, at a single fixed point\footnote{For higher derivatives acting on an object $X$, we will use the notation $\nabla_{\mu_1\cdots \mu_n}X:=\nabla_{\mu_1}\cdots \nabla_{\mu_n} X$.} $x'$,
 \begin{align}
 \begin{split}\label{eq:A_fock_series}
A_\mu(x'+x) &=
\sum_{n=0}^{\infty} \frac{1}{n! (n+2)} x^{\rho} x^{\mu_1}\cdots x^{\mu_n} \nabla_{\mu_1\cdots \mu_n}F_{\rho\mu}(x').
 \end{split}
\end{align}
In its turn, this expansion implies that the derivatives of the gauge field $A_\mu$, in the coincidence limit where $x'$ tends to $x$, can be expressed in terms of the derivatives of the field strength\footnote{For an expression $b(x,x')$ that depends on two coordinates, the definition is $[b(x,x')]:=\lim_{x'\to x} b(x,x')$. Although the gauge field $A_\mu$ and its derivatives do not depend explicitly on $x'$, they acquire an implicit dependence through the gauge fixing condition.}:
\begin{align}\label{eq:coincidenceA_Field}
[A^{\nu}{}_{;\mu_1\cdots\mu_n}(x)]=\frac{n}{n+1} F_{(\mu_1}{}^{\nu}{}_{;\mu_2\cdots \mu_n)}(x).
\end{align}
Note that we indicate idempotent symmetrization of indices by enclosing them in parentheses.

Coming back to the goal of obtaining a resummation valid for strong fields, a glimpse at Eq.~\eqref{eq:HK_eq_EM} gives us the following two intuitions and facts.
First, the term linear in spatial derivatives should be irrelevant for computing the diagonal of the heat kernel, since $A_\mu$ vanishes in the coincidence limit. \emph{A posteriori}, we will find that it is the antisymmetry of $F_{\mu\nu}$ that allows us to disregard this term altogether. Second, jafter discarding the linear term, the problem is reduced to the analysis of a Klein--Gordon operator with a nontrivial potential, as given in Eq.~\eqref{eq:VforEM}.

Furthermore, it should be clear that, due to the gauge fixing in Eq.~\eqref{eq:coincidenceA_Field} and the explicit form of the potential, contributions containing only powers of the field strength appear solely in the coincidence limit of the second derivative of $V_{EM}$. Indeed, the explicit expressions for its first derivatives read
\begin{align}\label{eq:gamma_EM}
    [V_{EM}] &=0,
    \\
    [\nabla_{\mu}V_{EM}]&=
    \tfrac{2}{3} \nabla_{\alpha }F^{ \alpha }{}_{\mu},
    \\
    [\nabla_{\mu\nu}V_{EM}]
    &=\frac{1}{2} F_{\mu\gamma } F^{\gamma}{}_{\nu} - \frac{1}{2} \nabla_{\gamma }\nabla_{(\mu}F^{\gamma }{}_{\nu)},
\end{align}
while higher derivatives of $V_{EM}$ will only yield terms involving invariants with one or more derivatives of the field strength. Considering the results from Ref.~\citenum{Brown:1975bc} and the insight that the linear derivative term in Eq.~\eqref{eq:operatorQ} is not relevant for our analysis, we can propose an Ansatz for the heat kernel that resum all the necessary contributions\footnote{Note that, contrary to the choice in Ref.~\citenum{Brown:1975bc}, we are working here in Euclidean space.}
\begin{align}\label{eq:HK_Y_ansatz}
 \begin{split}
  &K_{EM}(x,x';\tau)
  =:\frac{1}{(4\pi)^{d/2}}\frac{e^{-\tau V_{EM}(x')-\frac{1}{4}\tilde\sigma^\mu A^{-1}_{\mu\nu}(x';\tau)\tilde\sigma^{\nu} -C(x';\tau)}}{ \det^{1/2}\left(\tau^{-1} A(x'; \tau) \right)}\Omega_{EM}(x,x';\tau),
  \end{split}
  \end{align}
where we have introduced several quantities:
\begin{align}\label{eq:HK_Y_ansatz_dic}
\begin{split}
  \tilde\sigma_\mu(x,x'):&= \nabla_{\mu}\sigma(x,x')+ B_\mu(x';\tau),
  \\
  A_{\mu\nu}(x;\tau):&=\left[ \frac{1}{\gamma}\tanh(\gamma \tau)\right]_{\mu\nu},
  \\
  B_{\mu}(x;\tau):&= 2 \nabla^{\nu}V_{EM} \left[\gamma^{-2}\left( 1-\sech(\gamma \tau)\right)\right]_{\nu\mu},
  \\
  C(x,\tau):&=  \nabla^{\mu}V_{EM} \left[-\tau \gamma^{-2} +\gamma^{-3}\tanh(\gamma \tau)\right]_{\mu\nu} \nabla^{\nu}V_{EM}
  \\
  &\hu+\tfrac{1}{2}\left[\log\bigl(\cosh(\gamma \tau)\bigr)\right]^{\mu}{}_{\mu},
  \\
  \gamma^2_{\mu\nu}:&=2\nabla_{\mu\nu}V_{EM}.
 \end{split}
\end{align}
In flat space, Synge's worldfunction is given by~\cite{DeWitt:2003} $\sigma(x,x'):=(x-x')^2/2$. Additionally, tensors like $\gamma_{\mu\nu}$ can be treated as matrices with respect to their tensorial indices, so that $\left(\gamma^n\right)^{\mu}{}_{\nu}:= \gamma^{\mu}{}_{\mu_1}\cdots \gamma^{\mu_{n-1}}{}_{\nu}$.

For the sake of clarity and rigor, let us state our resummation claim more explicitly. Our assertion is that, when the function $\Omega_{EM}$ is expanded in powers of the proper time $\tau$,
\begin{align}\label{eq:HK_Y_expansion}
    \Omega_{EM} =:\sum_{j=0}^\infty a_j(x,x') \tau^{j-d/2},
\end{align}
the coincidence limit of the generalized Gilkey--Seeley--DeWitt (GSDW) coefficients, $[a_j]:=[a_j(x,x')]$, will be independent of the geometric invariants included in the set
\begin{align}\label{eq:invariants_Y}
    \mathcal{K}_{EM}=\lbrace F^{\mu}{}_{\nu_1} F^{\nu_1}{}_{\nu_2}\cdots F^{\nu_i}{}_{\mu}, \quad i\geq 0\rbrace.
\end{align}
The geometric invariants in $\mathcal{K}_{EM}$ are referred to as \emph{chains} for evident reasons. An important point is that we do not rule out the possibility that the field strength tensor may appear in the generalized GSDW coefficients; instead, in general it will appear but, in such instances, it will contribute to the formation of other geometric invariants, such as through contractions with derivatives of the field strength.

The proof of the resummation can be conducted via induction by examining a recursion relation for the GSDW coefficients. In effect, by replacing Eqs.~\eqref{eq:HK_Y_ansatz} and \eqref{eq:HK_Y_expansion} into the heat equation, we can straightforwardly derive the expression
\begin{align}
\begin{split}\label{eq:HK_Y_recurrence}
&-\left( j+1+\nabla_\alpha\sigma \nabla^\alpha \right) a_{j+1}(x,x')
= (-\nabla^2 +\mathfrak{S}) a_{j}(x,x')
\\
&\hspace{0.5cm}+ \sum_{n=1}^{\lfloor j/2 \rfloor} \frac{B_{2n}}{(2n)!}
\Big( 4(2^{2n}-1) \nabla^\alpha V' \left(\gamma'\right)_{\alpha\beta}^{2(n-1)}
+2^{2n} \nabla^\alpha\sigma \left(\gamma'\right)^{2n}_{\alpha\beta}\Big)
\nabla^\beta a_{j+1-2n}(x,x'),
\end{split}
\end{align}
where $B_n$ denote the Bernoulli numbers, $\lfloor \cdot \rfloor$ is the floor function, and we introduce what we call the effective potential
\begin{align}\label{eq:frakS}
\begin{split}
\mathfrak{S}(x,x';\tau):&=
  V(x) -  V(x')
-  \nabla^\alpha\sigma(x,x') {\nabla}_{\alpha }V(x')
-\frac{1}{4}  \nabla^{\alpha}\sigma(x,x')\nabla^\beta \sigma(x,x') (\gamma')^2_{\alpha \beta }.
\end{split}
\end{align}
Knowing that the initial condition in Eq.~\eqref{eq:HK_initial_condition} and the antisymmetry of the field strength fix the first GSDW coefficient to be
\begin{equation}\label{eq:a0}
a_0(x,x')=1\,,
\end{equation}
using Eq.~\eqref{eq:HK_Y_recurrence} one can determine the coincidence limit of a coefficient (or of its derivative) if the preceding coefficients and some of their derivatives are known. To be more concrete, the order implied by Eq.~\eqref{eq:HK_Y_recurrence} starts with $[a_0]$, $[\nabla_\alpha a_0]$, $[\nabla_{\alpha\beta} a_0]$, $[a_1]$, $\cdots$.

The essence of our results for an electromagnetic field can now be grasped from Eqs.~\eqref{eq:HK_Y_recurrence} and~\eqref{eq:frakS}. Notably, the Ansatz~\eqref{eq:HK_Y_ansatz} extends beyond the conventional Schwinger–DeWitt approach by effectively capturing local information about the potential and its derivatives. This ensures that the effective potential $\mathfrak{S}$ vanishes rapidly enough in the coincidence limit, thereby preventing the direct formation of chains in the prefactors that multiply the various GSDW coefficients (and their derivatives) in Eq.~\eqref{eq:HK_Y_recurrence}. Moreover, the tensorial structure of of Eq.~\eqref{eq:HK_Y_recurrence} is such that it precludes the indirect emergence of chains through the mixing of the derivatives of the GSDW coefficients with their prefactors. This assertion was proven inductively in Ref.~\citenum{Franchino-Vinas:2023wea} by introducing the concept of \textit{semi-chains}, which are formed by bisecting chains and leaving the indices free at the cut. For instance, from $F^{\mu}{}_{\nu}F^{\nu}{}_{\mu}$ we obtain the semi-chain $F^{\nu}{}_{\mu}$, with free indeces $\mu$ and $\nu$.

Overall, the proof holds for any dimension $d$ and applies to the local heat kernel, without requiring integrations by parts. In four dimensions, it is well-known that the chains can be expressed in terms of the invariants $\mathcal{F}$ and $\mathcal{G}$, which completes the proof of the Navarro-Salas--Pla conjecture.

On a more formal perspective, Eq.~\eqref{eq:HK_Y_recurrence} also indicates that, in our case, the GSDW coefficients contain only natural powers of the potential and its derivatives, given that it solely contains even powers of $\gamma_{\alpha\beta}$. This is in agreement with the general heat-kernel theory of operators of Laplace type.

Finally, considering the diagonal of our Ansatz in expression~\eqref{eq:HK_Y_ansatz},
\begin{align}\label{eq:HK_diagonal}
\begin{split}
&K_{EM}(x,x;\tau)
=\frac{e^{\tfrac{4}{9} \nabla_{\gamma}F^{\alpha \gamma} \left\lbrace \gamma^{-3} \left({\gamma \tau - 2 \tanh(\tfrac{1}{2} \gamma \tau)}\right)\right\rbrace_{\alpha \beta } \nabla_{\delta}F^{\beta \delta} }}
{(4\pi)^{\dime/2} \;{\det} ^{1/2}\big((\gamma \tau)^{-1} \sinh(\gamma \tau)\big) }  \Omega_{EM}(x,x;\tau),
\end{split}
\end{align}
we can readily obtain the Euclidean effective action by means of Eq.~\eqref{eq:effective_action}.
This expression extends beyond the Euler--Heisenberg action, as the exponential prefactor already incorporates derivatives of the field strength.
In any case, if one is willing to include derivatives, the first terms in the proper time expansion of $\Omega_{EM}$ provide additional insights, as detailed in Ref.~\citenum{Franchino-Vinas:2024wof}. By performing a Wick rotation, we arrive at the Minkowskian effective action~\footnote{For a complex scalar field, $n_s$ equals one in Eq.~\eqref{eq:effective_action}.},
\begin{align}
\Gamma_{\rm M}= \int \dxd[x][\dime] \int_0^\infty \frac{\dx[\tau]}{\tau}
\frac{e^{\tfrac{4}{9} \nabla_{\gamma}\tilde F^{\alpha \gamma} \left\lbrace \tilde\gamma_{\epsilon}^{-3} \left({\tilde\gamma_{\epsilon} \tau - 2 \tanh(\tfrac{1}{2} \tilde\gamma_{\epsilon} \tau)}\right)\right\rbrace_{\alpha \beta } \nabla_{\delta}\tilde F^{\beta \delta} }}
{(4\pi)^{\dime/2} \;{\det} ^{1/2}\big((\tilde\gamma_{\epsilon} \tau)^{-1} \sinh(\tilde\gamma_{\epsilon} \tau)\big) } \tilde\Omega_{EM}(x,x;\tau),
\end{align}
where the tilde denotes the Wick-rotated quantities. Additionally, we have introduced a parameter $\epsilon\to 0^+$, which parameterizes the Wick rotation and regulates the several poles that arise in the  Minkowskian setup. Importantly, the poles in the prefactor affect not only the $\tilde a_0$ term of $\tilde\Omega_{EM}$, from which one can reproduce the well-known Euler--Heisenberg Lagrangian, but also the higher coefficients. In other words, in this approach, the probability of pair creation for  a non-necessarily homogeneous electric field $E$ takes the form:
\begin{align}
    P=\int \dxd[x][\dime] \sum_{n=1}^\infty \left(b_0+b_1\cdots \right) e^{-\frac{n \pi m^2}{E}},
\end{align}
where $b_i$ represents a factor related to the $i$th GSDW coefficient and includes derivatives of the electric field.


\section{Resummation for a general vector coupled to linear derivatives}\label{sec:non-gauge}

The results discussed in the previous section lead us to the following question: Is this an isolated finding, depending on the specific properties of Abelian gauge fields, or can we derive a broader class of similar resummations, applicable to more general theories?

 We conjecture that the latter is indeed the case. To support this hypothesis, let us consider the following equation for the heat kernel:
\begin{align}
 \left\lbrace \partial_\tau+\left(-\nabla^2-2N^\alpha \nabla_\alpha+m^2\right)\right\rbrace K_N(x,x';\tau)=0.
\end{align}
This case is partially motivated by a simple model, consisting of a quantum massive, complex scalar field coupled to a vector background $N_\mu$ through a single derivative term.
While this theory bears some resemblance to the one described previously, it is crucial to note that the field $N^\mu$ is not a gauge field; it is not constrained by any gauge conditions or Bianchi identities, and there are no potential terms $V_{EM}$ involved. 
Moreover, this model can be viewed as a simplified representation of the coupling between fermions and torsion~\cite{Shapiro:2001rz}, as well as related axial potentials~\cite{McKeon:1998et,Copinger:2022bwl}, or, more ambitiously, non-Abelian gauge fields. Of course, one can consider a slightly generalized scenario in which the vector field $N_\mu$ is trivial in a vector bundle, namely $N_\mu = N_\mu \delta_{ab}$, with an immediate generalization of the upcoming results.

Inspired by the results in the previous section, we conjecture that the invariant $N^2$ can be resummed and propose the Ansatz
\begin{align}
    K_N(x,x';\tau) &=: e^{-\tau (N^2(x')+m^2)-N'_\mu\nabla^\mu \sigma(x,x')-\frac{\sigma(x,x')}{2\tau}} \Omega_N(x,x';\tau),
\end{align}
whose substitution in the heat equation leads to the following partial differential equation for $\Omega$:
\begin{align}
\begin{split}
&\Bigg\lbrace \frac{\dime}{2 \tau} + 2  (N-  {N'})^{\alpha }  \left( N' - \nabla\right)_{\alpha }
\\
&\hu\hu+ \nabla_{\alpha }\sigma (x,x') \frac{ (N- {N'}+\nabla)^\alpha}{\tau}- \nabla^2 + \partial_\tau
\Bigg\rbrace
\Omega_N (x,x';\tau)=0.
\end{split}
\end{align}
As customarily, we expand $\Omega_N$ in powers of the proper time $\tau$,
introducing the generalized GSDW coefficients $a_j^{(N)}$,
\begin{align}
    \Omega_N(x,x';\tau)&= \sum_{j=0}^{\infty} a_j^{(N)}(x,x') \tau^{j-n/2},
\end{align}
so that a recursion relation for them follows:
\begin{align}
\begin{split}
& \Big( j + \nabla_{\alpha }\sigma (x,x') (N-N'+\nabla)^{\alpha }  \Big)
a^{(N)}_{j}(x,x')
\\
&\hu\hu\hu=
 -2( N-  {N'})^{\alpha } \big(N' -\nabla\big)_{\alpha } a^{(N)}_{ j-1}(x,x')
 +\nabla^2 a^{(N)}_{ j-1}(x,x').
\end{split}
\end{align}
This expression is reminiscent of Eq.~\eqref{eq:HK_Y_recurrence} in that it features an ``effective vector coupling'' that vanishes in the coincidence limit. This property allows us to demonstrate that the geometric invariant $N^2$ does not appear in any of the generalized GSDW coefficients $a_j^{(N)}$, $j=0,\cdots$. In the terminology used in the previous section, $N^2$ constitutes the only chain, while the only available semi-chain is represented by the factor $N^\alpha$, where $\alpha$ is a free index. It is straightforward to show by induction that neither the coefficients nor their derivatives, in the coincidence limit, will generate this semi-chain, and consequently, no chain will arise in the generalized GSDW coefficients. In doing so, an important remark is that the derivatives of $a_0(x,x')$ are not trivial when the coincidence limit is taken.

A simple check that $N^2$ has indeed been resummed can be performed by directly computing the first improved GSDW coefficients, which read
\begin{align}
[a_1^{(N)}]&=- \nabla_{\alpha }N^{\alpha },\label{eq:a1}
\\
[a_2^{(N)}]&=
\tfrac{1}{6} \Big(3 (\nabla_\alpha N^\alpha )^2 -  \nabla_{\alpha }N_{\beta } \nabla^{\beta }N^{\alpha } -  \nabla_{\beta }N_{\alpha } \nabla^{\beta }N^{\alpha } - 2 N^{\alpha } \nabla^2N_{\alpha } -  \nabla_{\alpha }\nabla^2N^{\alpha }\Big),\label{eq:a2}
\\
\begin{split}
[a_3^{(N)}]&=
- \tfrac{1}{6} \nabla_{\alpha }N^{\alpha } \nabla_{\beta }N^{\beta } \nabla_{\gamma }N^{\gamma } + \tfrac{1}{3} N^{\alpha } N^{\beta } \nabla_{\gamma }N_{\beta } \nabla^{\gamma }N_{\alpha } + \tfrac{1}{6} \nabla_{\alpha }N^{\alpha } \nabla_{\beta }N_{\gamma } \nabla^{\gamma }N^{\beta }
\\
&\hu+ \tfrac{1}{6} \nabla_{\alpha }N^{\alpha } \nabla_{\gamma }N_{\beta } \nabla^{\gamma }N^{\beta } + \tfrac{1}{3} N^{\alpha } \nabla^{\beta }N_{\alpha } \nabla_{\beta \gamma }N^{\gamma } + \tfrac{4}{45} \nabla_{\alpha }{}^{\beta }N^{\alpha } \nabla_{\beta \gamma }N^{\gamma }
\\
&\hu-  \tfrac{2}{45} \nabla_{\alpha \gamma }N_{\beta } \nabla^{\beta \gamma }N^{\alpha } -  \tfrac{1}{45} \nabla_{\beta \gamma }N_{\alpha } \nabla^{\beta \gamma }N^{\alpha } + \tfrac{1}{3} N^{\alpha } \nabla_{\beta }N^{\beta } \nabla^2 N_{\alpha }
\\
&\hu-  \tfrac{1}{36} \nabla^2N^{\alpha } \nabla^2 N_{\alpha } -  \tfrac{1}{90} \nabla_{\alpha }{}^{\beta }N^{\alpha } \nabla^2 N_{\beta }
-  \tfrac{1}{15} \nabla^{\beta }N^{\alpha } \nabla_{\alpha }\nabla^2N_{\beta }
\\
&\hu -  \tfrac{1}{15} \nabla^{\beta }N^{\alpha } \nabla_{\beta }\nabla^2N_{\alpha } + \tfrac{1}{6} \nabla_{\alpha }N^{\alpha } \nabla_{\beta }\nabla^2N^{\beta } -  \tfrac{1}{30} N^{\alpha } \nabla^4N_{\alpha }
\\
&\hu -  \tfrac{1}{60} \nabla_{\alpha }\nabla^4N^{\alpha }.\label{eq:a3}
\end{split}
\end{align}
In line with our claim, there is no contribution proportional to $N^2$ in these coefficients; however, geometric scalars formed by contractions of $N^\alpha$ with derivative contributions are indeed present. As a final note, it is important to highlight that the coefficients in Eqs.~\eqref{eq:a1}-\eqref{eq:a3} have been derived without using integration by parts, and no boundary contributions have been neglected. Therefore, these coefficients can be employed to compute local quantities and can be effectively combined with smearing functions.


\section{Outlook}\label{sec:outlook}

We have shown some powerful aspects of the heat kernel techniques in obtaining resummed expressions for effective actions in background fields. For an electromagnetic background field, the Navarro-Salas--Pla conjecture~\cite{Navarro-Salas:2020oew}, namely that a resummation of the invariants $F_{\mu\nu}F^{\mu\nu}$ and $F_{\mu\nu}\tilde F^{\mu\nu}$ was available for $\dime=4$, has been extended and reviewed (see  Ref.~\citenum{Franchino-Vinas:2023wea} for a more detailed proof).

In addition, we conjecture that analogous resummations may exist for more general backgrounds. While this manuscript focuses on a simple toy model where the background is coupled with the first derivative of the quantum field, there are numerous potential generalizations, including nontrivial gravitational backgrounds. These have clear cosmological implications, ranging from pair creation in de Sitter space~\cite{Bavarsad:2017oyv} to broader claims regarding generalized Hawking radiation~\cite{Wondrak:2023zdi,Ferreiro:2023jfs,Hertzberg:2023xve,Akhmedov:2024axn}, as well as condensed matter systems, in which fermions might couple to the axial vector part of the torsion~\cite{deJuan:2009ldt,Ferreiros:2018udw,Vozmediano:2010zz,Baggioli:2020gpf,Roy_2018}.

A further promising alternative is to explore the use of the covariant derivative expansion, a technique that shares some features with the heat kernel and has already demonstrated its effectiveness in addressing gravitational setups~\cite{Larue:2023uyv}.


\section*{Acknowledgments}
SAF thanks the members of LAPTh, Annecy, especially J.~Quevillon, for their warm hospitality. SAF acknowledges the support from Helmholtz-Zentrum Dresden-Rossendorf. SAF and FDM acknowledge the support from Consejo Nacional de Investigaciones Científicas y Técnicas (CONICET) through Project PIP 11220200101426CO.
SAF and UWH acknowledge the support of UNLP through Project 11/X748. The work of SAF and VV has been partially funded by Next Generation EU through
the project ``Geometrical and Topological effects on Quantum Matter (GeTOnQuaM)''. The research activities of SAF, CGP and VV have been carried out in the framework of the INFN Research Project QGSKY. The Authors extend their appreciation to the Italian National Group of Mathematical Physics (GNFM, INdAM) for its support. The authors would like to acknowledge networking support by the COST Action CA18108.


\bibliographystyle{ws-procs961x669}
\bibliography{bibliografia}

\end{document}